\documentclass[journal]{IEEEtran}

\usepackage{graphicx}
\usepackage{lineno}
\usepackage{stfloats}
\usepackage{xcolor}
\usepackage{ragged2e}

\ifCLASSINFOpdf

\else

\fi

\begin{document}

\title{Compact free-running InGaAs/InP single-photon detector with 40\% detection efficiency and 2.3 kcps dark count rate}

\author{Qi~Xu, 
        Chao~Yu, 
        Wei~Chen, 
		Jianglin~Zhao, 
		Dajian~Cui, 
		Jun~Zhang, and 
		Jian-Wei~Pan
\thanks{Qi Xu and Chao Yu contributed equally to this work. They are with Hefei National Research Center for Physical Sciences at the Microscale and School of Physical Sciences, University of Science and Technology of China, Hefei 230026, China, and are also with CAS Center for Excellence in Quantum Information and Quantum Physics, University of Science and Technology of China, Hefei 230026, China.}
\thanks{Wei Chen, Jianglin Zhao and Dajian Cui are with Chongqing Optoelectronics Research Institute, Chongqing 400060, China, and are also with Chongqing Key Laboratory of Core Optoelectronic Devices for Quantum Communication, Chongqing 400060, China (e-mail: cuidj@cetccq.com.cn).}
\thanks{Jun Zhang and Jian-Wei Pan are with Hefei National Research Center for Physical Sciences at the Microscale and School of Physical Sciences, University of Science and Technology of China, Hefei 230026, China, and with CAS Center for Excellence in Quantum Information and Quantum Physics, University of Science and Technology of China, Hefei 230026, China, and also with Hefei National Laboratory, University of Science and Technology of China, Hefei 230088, China (e-mail: zhangjun@ustc.edu.cn).}}
\maketitle

\begin{abstract}
Free-running InGaAs/InP single-photon detectors (SPDs) based on negative-feedback avalanche diodes (NFADs) are the key components for applications requiring asynchronous single-photon detection in the near-infrared region. From the perspective of practical applications, the features of SPDs in terms of high photon detection efficiency (PDE), low noise, large sensitive area, and compactness are highly desired for system integration and performance enhancement. Here, we present the implementation of a compact four-channel multimode fiber coupling free-running InGaAs/InP SPD, with the best overall performance to date. On the one hand, we design and fabricate structure-optimized InGaAs/InP NFAD devices with 25 $\mu$m diameter active area and integrated thin film resistors to enhance the maximum achievable PDE. On the other hand, we apply a compact thermoacoustic cryocooler to regulate the operating temperature of NFADs within a large range, and design a dedicated readout circuit with minimized parasitic parameters and tunable settings of hold-off time to suppress the afterpulsing effect. The SPD is then characterized to achieve remarkable overall performance simultaneously at 1550 nm, i.e., 40\% PDE, 2.3 kcps dark count rate, 8\% afterpulse probability and 49 ps timing jitter (full width at half maximum) under the conditions of 5.9 V excess bias voltage, 10 $\mu$s hold-off time and 213 K operation temperature. Such performance and the results of the long-term stability tests indicate that the SPD could be a favorable solution for practical applications.
\end{abstract}


\section{Introduction}

\IEEEPARstart{S}{ingle-photon} detectors (SPDs)~\cite{SPD11} are the key components in numerous applications such as quantum key distribution (QKD), light detection and ranging, optical time-domain reflectometry, and fluorescence lifetime imaging microscopy. In the near-infrared range, InGaAs/InP SPDs are the primary candidate for practical applications due to the advantages of compactness, low cost, and ease of use~\cite{JMH15}. InGaAs/InP SPDs usually consist of InGaAs/InP single-photon avalanche diodes (SPADs), readout circuits and affiliated circuits. Such SPDs can be operated either in gating mode or in free-running mode. The gating mode is suited for synchronous single-photon detection in which photons arrive periodically. In contrast, the free-running mode is suited for the scenario where the arrival time of the photon is unknown.

\begin{figure*}[t]
\centering
\includegraphics[width=18 cm]{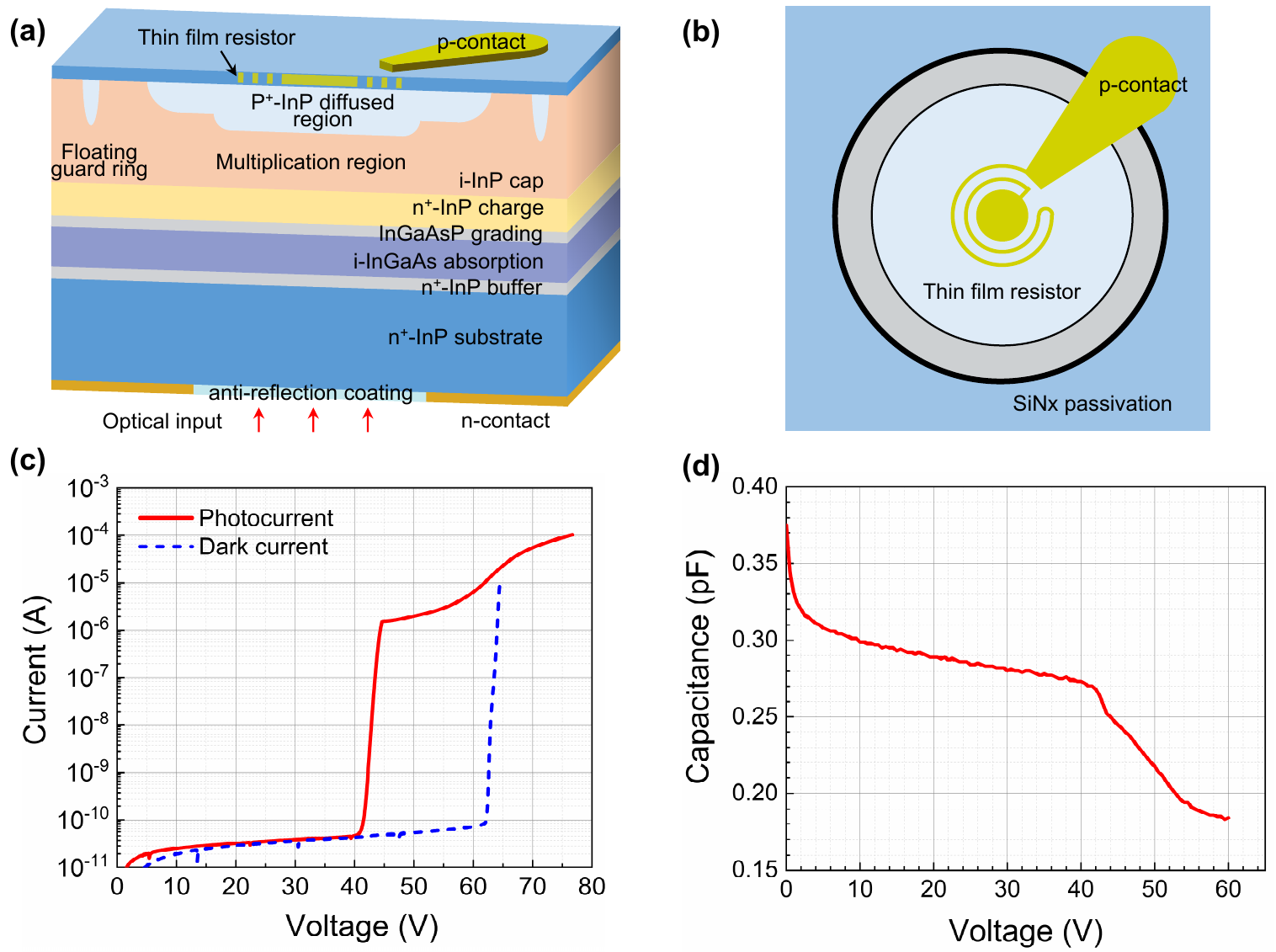}
\caption{(a) Schematic diagram of the NFAD device with an SAGCM SPAD and a thin film resistor. (b) Top view of the thin film resistor. (c) Measured I-V curve of the NFAD device. (d) Measured C-V curve of the NFAD device.}
\label{fig1}
\end{figure*}

To date, several techniques have been reported to implement free-running InGaAs/InP SPDs~\cite{JMH15}. Passive quenching is a fundamental approach for free-running mode SPD~\cite{JTK00,MCJ08,RMG09,CMX10}. However, the discrete components introduce a large stray capacitance, leading to excessive charge flow through a SPAD, which results in a severe afterpulsing effect. Using the active quenching technique can rapidly quench avalanches, and therefore suppressing the afterpulsing effect~\cite{RDJ07,JRJ09}, but it is difficult to implement an extremely short quenching time at the subnanosecond level. Recently, the quenching time has been reported as low as 0.7 ns for Silicon SPAD~\cite{Xu20}. In the past decade, the primary approach for implementing practical free-running InGaAs/InP SPDs has been using negative-feedback avalanche diodes (NFADs)~\cite{XMB09,MXB10,XMK11}. NFAD devices monolithically integrate high-resistance thin
film resistors on the surface of dies. Due to the monolithic integration, the parasitic capacitance of the quenching circuit is minimized, and along with the setting of an appropriate active hold-off time, the afterpulsing effect of the NFAD devices can be significantly suppressed.

Typical parameters for characterizing InGaAs/InP SPDs include photon detection efficiency (PDE), dark count rate (DCR), afterpulse probability ($P_{ap}$), timing jitter, and maximum count rate~\cite{JMH15}. These parameters may affect each other, which implies that simply improving one parameter may induce the deterioration of other parameters. For instance, raising the excess bias voltage can effectively increase the PDE and reduce the timing jitter, but this can lead to increases in the DCR and $P_{ap}$. Cooling down NFAD devices can reduce DCR, but this can also result in an increase in $P_{ap}$ and a decrease in photon absorption efficiency~\cite{FMA13}. Setting a long hold-off time can suppress $P_{ap}$, but the maximum count rate is thus limited. As a result, all the parameters of InGaAs/InP SPDs have to be compromised, and therefore continuously improving the overall performance of SPDs is crucial for practical applications. For gating mode, we previously reported a 1.25 GHz gating InGaAs/InP SPD with a record performance of 60\% PDE and 340 kcps DCR as well as a practical performance of 40\% PDE and 3 kcps DCR~\cite{YWT20}. Moreover, several groups have also reported InGaAs/InP SPADs with over 50\% PDE~\cite{Tada20,Tosi21,He22}. For the free-running mode, the overall performance of SPD has been improving during the past decade~\cite{ZDA12,TCO12,BNT14,CMH17,CJH18}.

In this paper, we present a compact four-channel free-running InGaAs/InP SPD with the best overall performance to date. We fabricate InGaAs/InP high-performance SPAD chips first and then monolithically integrate thin film resistors on the surface to implement NFAD devices. The NFAD device is packaged inside a transistor-outline-46 (TO-46) shell with either multimode fiber (MMF) coupling using an aspheric microlens or single-mode fiber (SMF) coupling using a spheric microlens. A thermoacoustic cryocooler is used to regulate the operating temperature of four NFADs, and a dedicated readout circuit with tunable settings of active hold-off time is designed to suppress the afterpulsing effect. The SPD is assembled inside a cabinet with a size of 21 cm$\times$28 cm$\times$9 cm. The free-running InGaAs/InP SPD is then characterized to simultaneously achieve 40\% PDE, 2.3 kcps dark count rate, 8\% afterpulse probability and 49 ps full width at half maximum (FWHM) timing jitter under the conditions of 5.9 V excess bias voltage, 10 $\mu$s hold-off time and 213 K operation temperature.

\section{NFAD DESIGN AND FABRICATION}

The NFAD device is composed of an InGaAs/InP SPAD with a monolithically integrated thin film resistor, as illustrated in Fig.~\ref{fig1}(a). We follow the method~\cite{JBL16} to design the InGaAs/InP SPAD using the structure of separate absorption, grading, charge, and multiplication (SAGCM) layers with a stepped PN junction and a floating guard ring. The electric field distributions in the InGaAs absorption layer and the InP multiplication layer are regulated by tuning the doping concentration in the grading and charge layers. In Geiger mode, the electric field strength in the absorption layer is regulated to $\sim$ 1.5$\times10^{5}$ V/cm to guarantee the drift rate of the photogenerated carriers reaching the saturation level and to avoid a significant tunneling effect. The thickness of the multiplication layer is designed in the range of 1.2$\sim$1.3 $\mu$m to provide sufficiently high electric field strength for obtaining a desired avalanche probability. A ladder structure is formed for the PN junction using the double-diffusion process technology. Such a design can create a uniform electric field in the central zone of the PN junction while reducing the high electric field regions at the edge of the PN junction, which can suppress the afterpulsing effect~\cite{YWT20,Jiang18}. Additionally, a floating guard ring is designed to reduce the edge effect of the electric field and to avoid the edge prebreakdown caused by the high electric field at the edge of the PN junction. Considering the factors of fiber coupling efficiency and DCR performance, the diameter of active area is designed to be 25 $\mu$m.

On the surface of the diffusion region, a thin film NiCr resistor meander line is monolithically integrated, as shown in Fig.~\ref{fig1}(b). The shape and position of the thin film resistor are carefully adjusted to minimize the parasitic capacitance, which can reduce both the quenching time and the afterpulsing effect. For the value choice of the resistor, multiple factors are considered such as the quenching time of NFAD, the recovery time of NFAD, and the process complexity of fabrication. In our NFAD devices, the thin film resistor value is designed as $\sim$ 200 k$\Omega$ by regulating the length of the meander line, which brings out a considerably short quenching time of $\sim$ 800 ps and a moderate recovery time of $\sim$ 150 ns.

\begin{figure*}[htbp]
\centering
\includegraphics[width=18 cm]{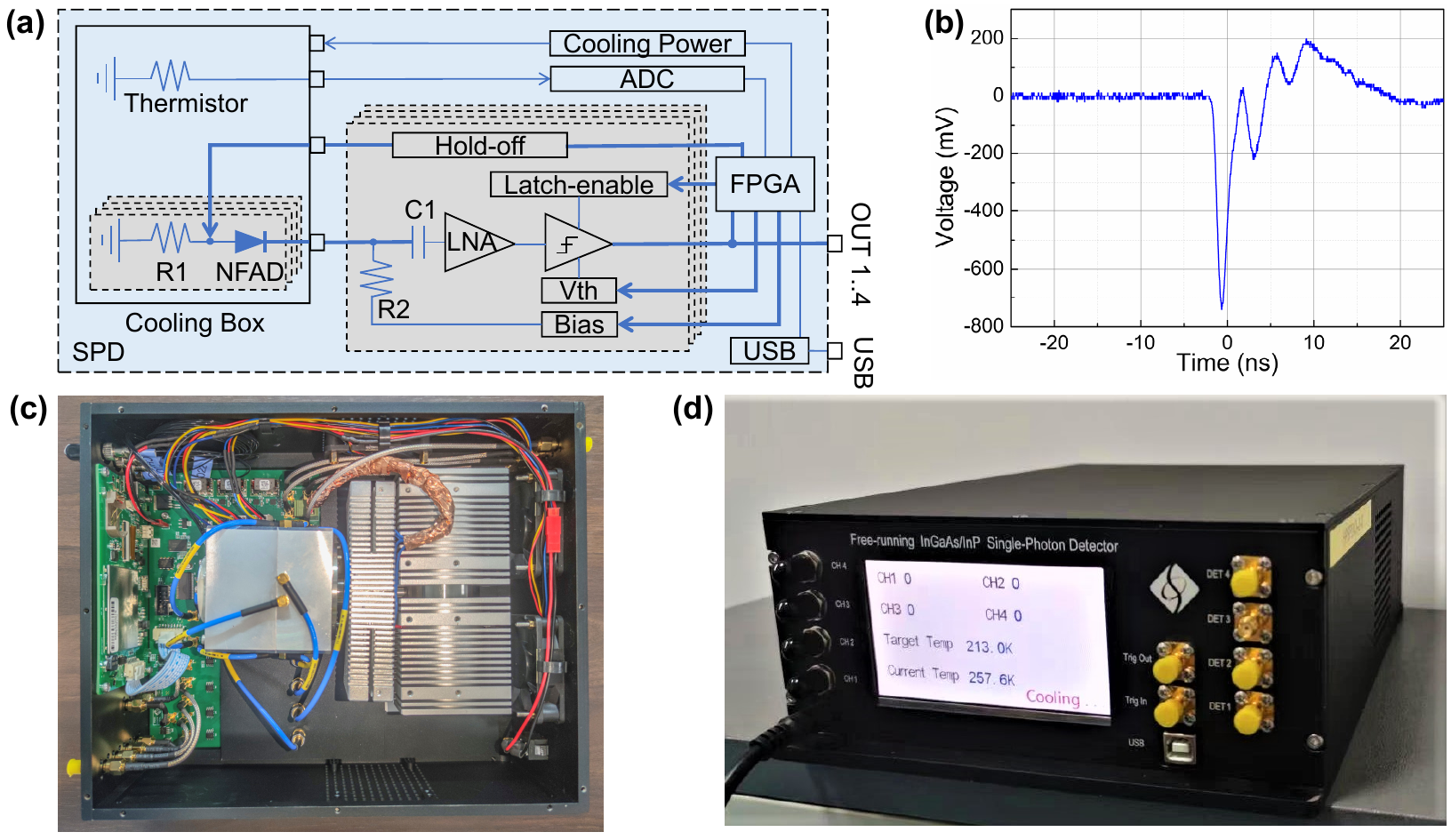}
\caption{(a) The design diagram of four-channel MMF coupling free-running InGaAs/InP SPD. LNA: low-noise amplifier. ADC: analog-to-digital converter. Vth: threshold voltage. USB: universal serial bus. (b) Typical avalanche signal at the output of the LNA, captured by an oscilloscope with 8 GHz bandwidth. (c) The structure photo inside the SPD. (d) The appearance photo of the SPD.}
\label{fig2}
\end{figure*}

The SPAD chip is fabricated via the standard epitaxial process of metal-organic chemical vapor deposition~\cite{YWT20}. During fabrication, the background impurity concentration of epitaxial materials is well regulated to a level as low as 1$\times$$10^{15}$ $cm^{-3}$, and the surface charge density in the charge layer is controlled with an accuracy of 5\%. The stepped PN junction structure is fabricated via the double-diffusion process technology using a Zn source with a depth accuracy of 50 nm~\cite{YWT20}. The thin film resistor is fabricated following magnetic control evaporation and etching processes. The line width of the resistor is controlled within an accuracy of 50 nm, and the square resistance is $\sim$ 1 k$\Omega$/sq.

The NFAD device is then characterized at room temperature before packaging. Figure~\ref{fig1}(c) plots current-voltage (I-V) curves with and without light illumination, from which one can observe that the punch-through voltage and the breakdown ($V_{br}$) voltage of the NFAD device are $\sim$ 40 V and 62 V, respectively. With a reverse bias voltage of $V_{br}$-1 V, the total dark current of the NFAD is measured to be an ultralow level of 0.1 nA, which is primarily contributed by the surface dark current. Given that the bias voltage is larger than the punch-through voltage, the photocurrent increases rapidly first and then slowly  increases due to the voltage drop on the integrated thin film resistor. Figure~\ref{fig1}(d) plots the capacitance-voltage (C-V) curve of the NFAD, which indicates that the capacitance of the NFAD device is $\sim$ 0.18 pF in Geiger mode.

\begin{figure}[htbp]
\centerline{\includegraphics[width=8 cm]{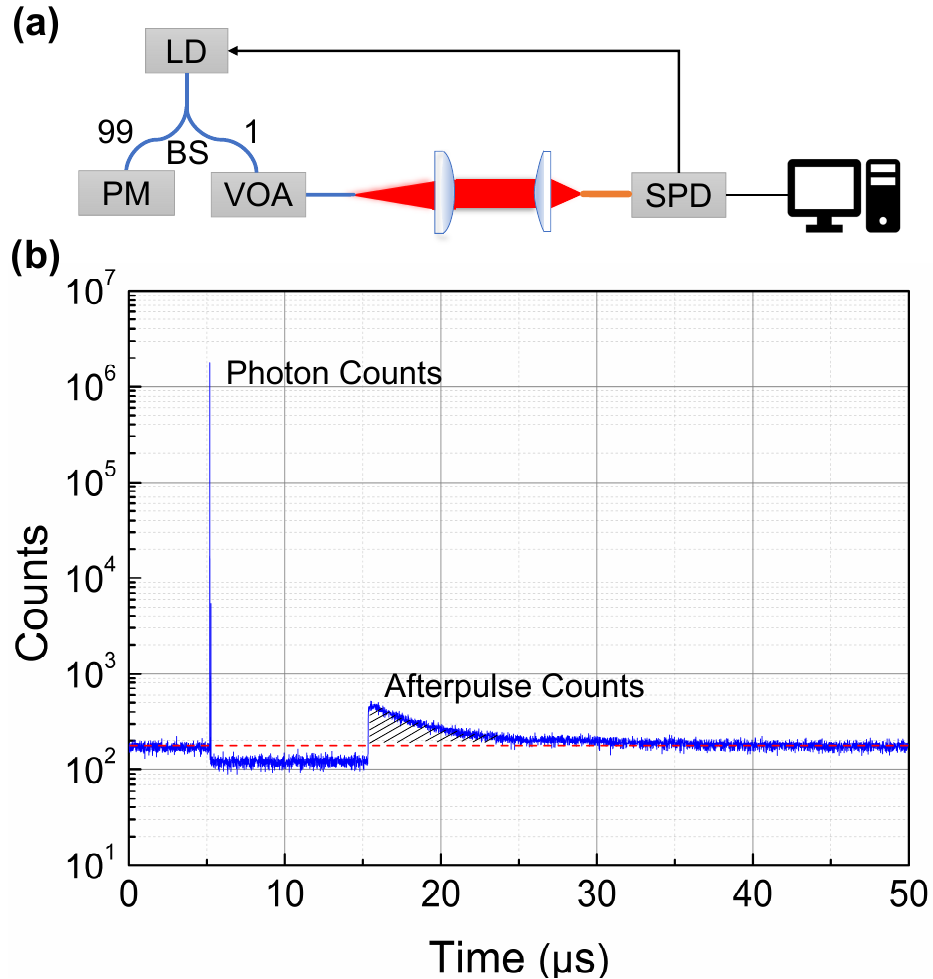}}
\caption{(a) Experimental setup for the free-running InGaAs/InP SPD characterization. LD: laser diode. PM: power meter. VOA: variable optical attenuator. (b) A typical detection event distribution measured by TDC with an acquisition time of 5 minutes under the condition of a 10 $\mu$s hold-off time, 5.9 V excess bias voltage and 213 K operation temperature.}
\label{fig3}
\end{figure}

\section{SPD SYSTEM AND CHARACTERIZATION}

In the experiment, the NFAD devices are encapsulated using standard TO-46 packages with MMF coupling. Inside the package, the incident photons emitted from the MMF with a 62.5 $\mu$m core diameter are focused on the photosensitive area of NFAD devices with a 25 $\mu$m diameter via an aspheric lens. The coupling efficiency is measured to be $\sim$ 90\% at 1550 nm. To optimize the overall performance of NFAD devices over a wide temperature range, a thermoacoustic cryocooler is used to regulate the operation temperature, which has the advantages of compactness, high reliability and excellent cooling power~\cite{VRG11}. The operation temperature of the NFAD devices is regulated by a proportional-integral-derivative program in a field programmable gate array (FPGA) within a range from 173 K to 273 K.

The design diagram of the free-running InGaAs/InP SPD is illustrated in Fig.~\ref{fig2} (a). Avalanche signals are capacitively coupled from the NFAD cathode, which are then amplified by a 40 dB low noise amplifier. Figure~\ref{fig2} (b) shows a typical amplified avalanche signal, exhibiting an amplitude of $\sim$ 750 mV, a rising time of 1.4 ns, and a time of full width at tenth maximum of $\sim$ 3.6 ns in the main peak. The avalanche signals are then discriminated to output LVTTL signals. The LVTTL signals are connected with the FPGA to generate hold-off signals with tunable time duration, during which the NFAD device is maintained in linear mode. Due to the capacitive responses, two signal spikes occur corresponding to the rising and falling edges of the hold-off signals. To avoid the detection of signal spikes, the discriminator is latched by the FPGA until an additional 30 ns after the end of the hold-off signals. The SPD provides a peripheral interface of a universal serial bus port to communicate with a personal computer. A LabView program is developed for the parameter setting of SPD, including bias voltage, hold-off time, threshold voltage, and operation temperature. The SPD is finally integrated into a compact module with a size of 21 cm $\times$ 28 cm $\times$ 9 cm, and its total weight is 4.5 kg. Fig.~\ref{fig2} (c) and Fig.~\ref{fig2} (d) show the layout structure inside the SPD and appearance of the SPD, respectively.

The SPD is then characterized following the standard calibration scheme~\cite{JMH15}, as illustrated in Fig.~\ref{fig3} (a). The SPD provides a 20 kHz signal to trigger a laser diode (LD). The laser pulses with a width of 70 ps are coupled into a single-mode fiber, and then divided by a 99:1 fiber beam splitter. The light intensity in the 99\% port is monitored by a power meter (PM), while the pulses from the 1\% port pass a variable optical attenuator (VOA) to further attenuate the intensity down to the single-photon level. To match the MMF coupling method, an SMF-to-MMF conversion module is used. In the module, the attenuated laser pulses are transmitted from the SMF to free space by a collimator and then coupled to an MMF by an aspheric lens. The overall coupling efficiency of the conversion module reaches 87.5\%. The detection signals of SPD are measured by a built-in time-to-digital converter (TDC) with a bin size of 10 ns implemented by the FPGA. A typical distribution histogram of detection events with an acquisition time of 5 minutes is listed in Fig.~\ref{fig3} (b). The main peak corresponds to the photon detection events, while the counts above the level of dark counts inside the second peak correspond to the afterpulse events.

\begin{figure}[htbp]
\centerline{\includegraphics[width=8 cm]{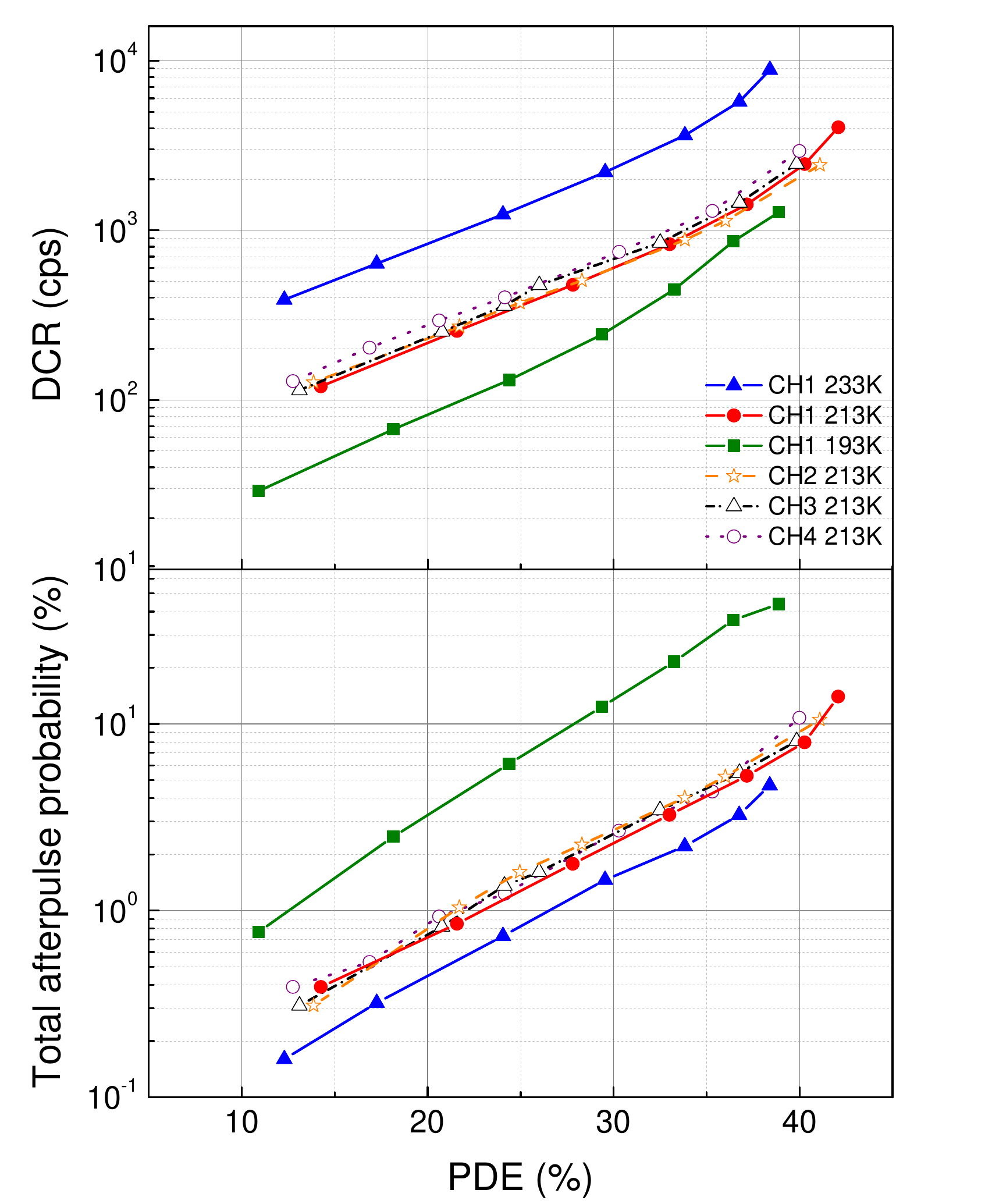}}
\caption{Dark count rate and total afterpulse probability versus PDE for 4 channels. The hold-off time is set to 10 $\mu$s. Solid lines plot the performance of CH1 SPD at 3 different temperature.}
\label{fig4}
\end{figure}

The PDE and $P_{ap}$ of the SPD can be calculated from the TDC data. Considering the Poisson distribution of the laser pulses, the PDE is calculated as~\cite{JMH15}
\begin{equation}
\label{PDE}
PDE=-\frac{1}{\mu}\ln(1-\frac{S_{p}}{f}),
\end{equation}
where $\mu$ is the mean photon number per laser pulse, $S_{p}$ is the detected photon count per second and $f$ is the repetition frequency of the pulsed laser. The afterpulse count rate $S_{ap}$ is obtained from the TDC data, and the total afterpulse probability is calculated as $P_{ap}$=$S_{ap}$/$S_{p}$.

Figure~\ref{fig4} plots DCR and $P_{ap}$ as a function of PDE at different temperatures with a hold-off time of 10 $\mu$s for 4 channels, from which one can observe that the 4 channels of SPDs exhibit very similar performances at 213 K. Therefore, for the following characterization we only select the results of CH1 SPD for presentation. From Fig.~\ref{fig4} (a), one can observe that in the region of the PDE below $\sim$ 35\% the trends of PDE versus DCR are basically exponential, while in the region of ultrahigh PDE, the increase in DCR is significantly faster than PDE. This is likely due to the nonlinear effect of the hold-off time setting, particularly in the case of a high count rate~\cite{CJH18}. Such an effect influences the maximum achievable PDE of SPD. As plotted in Fig.~\ref{fig4}, the maximum achievable PDE at 213 K is higher than that at the other two temperatures. At 213 K, the maximum PDE of SPD reaches 42.1\% with 3.7 kcps DCR and 14\% $P_{ap}$. For practical uses of SPD, 2 kcps DCR could be a suitable standard for comparison. Given this value as a reference, at 213 K PDE slightly dropped to 40\% while DCR and $P_{ap}$ considerably decreased to 2.3 kcps and 8\%, respectively. To the best of our knowledge, such overall performance could be the best to date concerning the free-running InGaAs/InP SPD.

\begin{figure}[htbp]
\centerline{\includegraphics[width=8 cm]{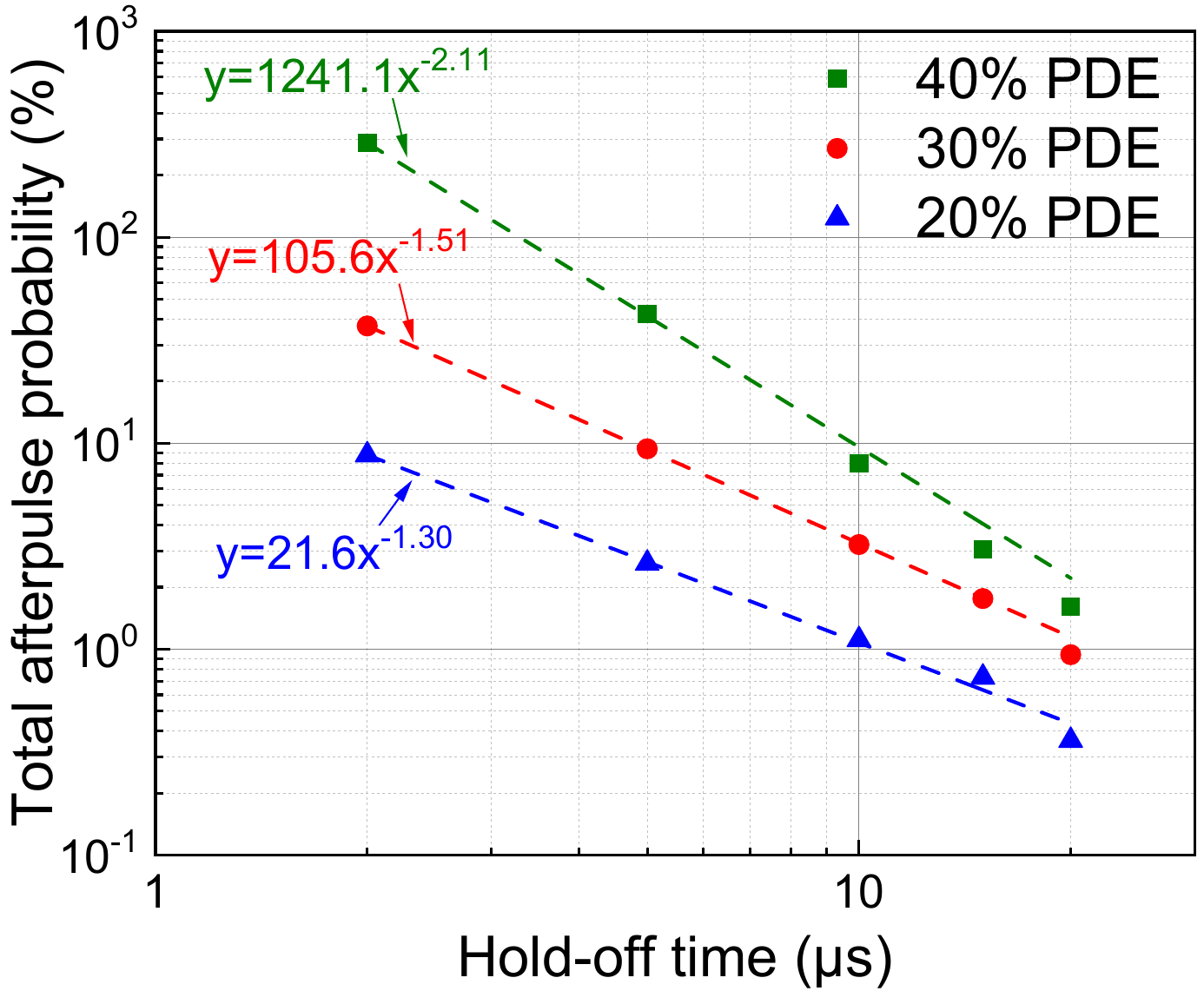}}
\caption{Afterpulse probability as a function of hold-off time in the cases of different PDE. The operation temperature of the NFAD device is 213 K. The data are fitted by a power-law.}
\label{fig5}
\end{figure}

We then investigate the relationship between $P_{ap}$ and hold-off time. Figure~\ref{fig5} plots $P_{ap}$ versus hold-off time with different PDE settings at 213 K. Intuitively, one might consider that $P_{ap}$ exponentially decays with increasing hold-off time since the lifetime of trapped carriers is highly related to hold-off time. However, previous experiments have already demonstrated that the temporal evolution of $P_{ap}$ might follow a simple power-law dependency, i.e., $P_{ap}$ $\propto$ $T_{ho}^{-\alpha}$ ~\cite{TCO12,BNT14,CJB11}, where $T_{ho}$ and $\alpha$ are the hold-off time and exponential value, respectively. This phenomenon has been explained by a dense spectrum of trap levels in the InP multiplication layer~\cite{MXM12,KLK15}.

In the experiment, the measured results are fitted by a power-law, see the dashed lines in Fig.~\ref{fig5}, and the fitting $\alpha$ values are 1.30, 1.51 and 2.11 in the cases of 20\% PDE, 30\% PDE and 40\% PDE, respectively. The relatively large $\alpha$ value in the case of 40\% PDE might be attributed to the high tunneling probability due to the strong electric field. The large $P_{ap}$ value at high PDE is due to the increasing number of avalanche carriers. In such a case, $P_{ap}$ is more sensitive to the setting of hold-off time than the cases of low PDE. For instance, in the case of 40\%
, $P_{ap}$ reaches $\sim$ 300\% with a 2 $\mu$s hold-off time but drastically drops down to 1.6\% with a 20 $\mu$s hold-off time. For comparison, in the case of 20\% PDE, $P_{ap}$ is measured to be $\sim$9\% with a 2 $\mu$s hold-off time while reduces to 0.4\% with a 20 $\mu$s hold-off time. Therefore, for applications requiring a high count rate, the SPD parameters, including PDE, hold-off time and operation temperature, must be comprehensively optimized.

\begin{figure}[htbp]
\centerline{\includegraphics[width=8 cm]{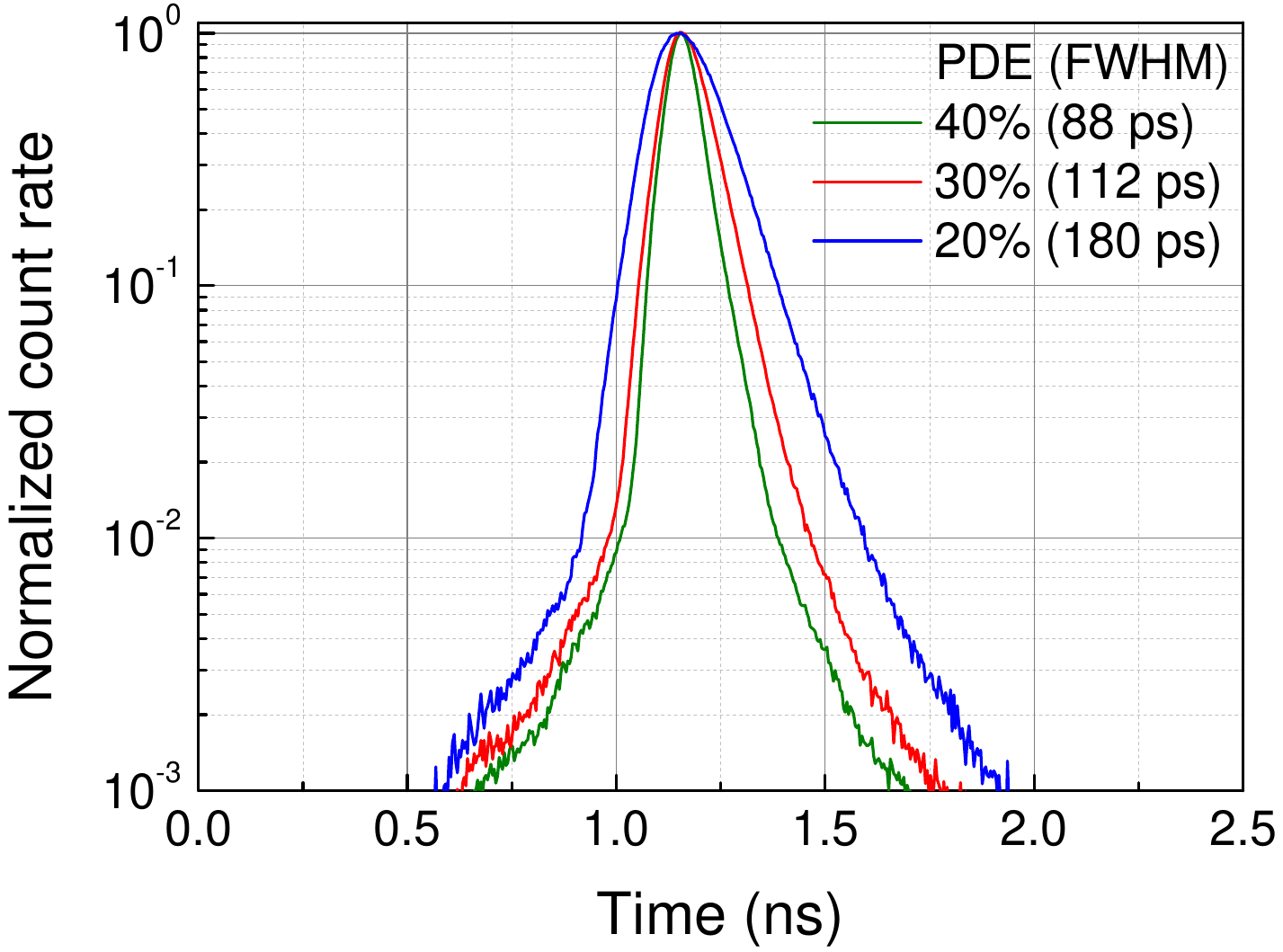}}
\caption{Timing jitter measurements of the free-running InGaAs/InP SPD at 213 K. The laser pulse width and the jitter of measurement electronics are measured to be 70 ps and 22 ps, respectively.}
\label{fig6}
\end{figure}

\begin{figure}[bp]
\centerline{\includegraphics[width=8 cm]{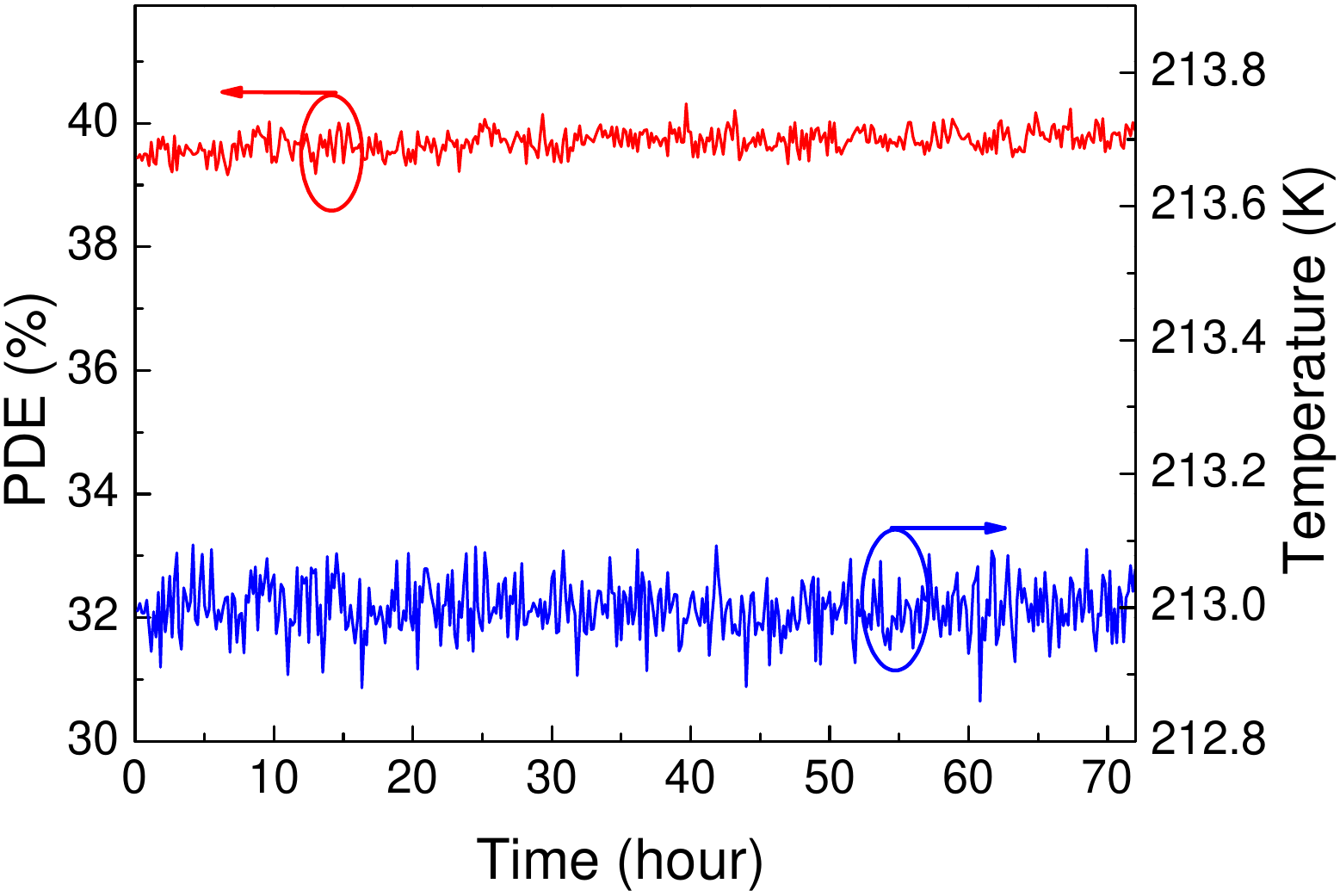}}
\caption{PDE and temperature stability tests for the free-running InGaAs/InP SPD.}
\label{fig7}
\end{figure}

Timing jitter is also an important parameter for the free-running SPD, which is primarily contributed by two factors~\cite{EGB16}. One is the transit time of the photogenerated holes from the InGaAs absorption layer to the InP multiplication layer. The other is the build-up time during the avalanche process. As the electric field strength in the InP multiplication layer increases, the build-up time of the avalanche and thus the corresponding time uncertainty could be highly shortened.
Figure~\ref{fig6} shows the total FWHM timing jitter measurements of the free-running InGaAs/InP SPD at 213 K. As the PDE increases from 20\% to 40\%, the total FWHM timing jitter significantly decreases from 180 ps to 88 ps. The total timing jitter includes the contributions by SPD, laser pulse width, jitter of laser pulses, and jitter of measurement electronics. The laser pulse width is measured to be 70 ps FWHM, while the jitter of laser pulses is negligible. The jitter of measurement electronics, including the signal generator and the external high resolution TDC system, is measured to be 22 ps FWHM.
Then, the intrinsic FWHM timing jitter of the free-running SPD is calculated to be 164 ps and 49 ps in the cases of 20\% PDE and 40\% PDE, respectively.

\begin{table*}[htbp]
\centering
\caption{Performance comparison of free-running InGaAs/InP SPD prototypes and products.}
\label{t1}
\renewcommand\arraystretch{2}
\begin{tabular}{lcccccc}
\hline
\hline
Ref                             & Temperature (K)     & PDE @1550 nm        & DCR (cps)        & Hold-off time ($\mu$s)   & Afterpulse probability   & Timing jitter (ps)   \\
\hline
This work                       & 213                 & 40\%              & 2.3 k            & 10                       & 8\%                      & 49                   \\
Ref~\cite{ZDA12}           & 193                 & 10\%                & 100              & /                       & /                        & 30                   \\
Ref~\cite{BNT14}        & 163                 & 27.7\%              & 15.2             & 20                       & 20\%                     & 129                  \\
Ref~\cite{JYY21}          & 223                 & 25\%                & 2 k              & 5                        & 4\%                      & /                    \\
QCD600B~\cite{QCD600B}          & /                   & 35\%                & 6 k              & 5                        & 8\%                      & 150                  \\
ID Qube~\cite{idqube}     & 223               & 25\%                   & 6 k                &/                      & /                    & 150                \\
PDM-IR~\cite{pdmir}             & 225                 & 22\%                & 6 k              & 20                       & /                        & 71                   \\
SPD\_A\_NIR~\cite{aurea}        & /                   & 10\%                & 1 k              & 10                       & 0.1\%                    & 180                  \\
Wooriro NFAD~\cite{Wooriro}      & 233                & 10\%                 & 10 k                   & /                         & /                       & /                  \\
\hline
\hline
\end{tabular}
\begin{flushleft}
\ \ \ \ \ \ For each prototype/product, the listed parameters can be achieved simultaneously.
\end{flushleft}
\end{table*}

We then perform 72-hour stability tests of the PDE and temperature for the free-running InGaAs/InP SPD. During the tests, the values of PDE and temperature are recorded every 10 minutes, as plotted in Fig.~\ref{fig7}. The target values of the PDE and operation temperature for the SPD are set as 40\% and 213.00 K, respectively. The measured temperature fluctuations reach a level as low as 0.04 K, as shown in Fig.~\ref{fig7}. Due to the high stability features of temperature and bias voltage, the mean value of PDE is 39.70\% with fluctuations of 0.19\% during the tests.

Finally, we perform a performance comparison between our SPD and other reported prototypes and commercial products of free-running InGaAs/InP SPD, as listed in Table~\ref{t1}, in which our SPD clearly exhibits significant improvements in overall performance. Considering the feature of high stability and the advantage of overall performance, we believe that the free-running SPD in this work could be a favorable solution for practical uses.

\section{Conclusion}

In conclusion, we have presented a compact four-channel free-running InGaAs/InP SPD with the best overall performance to date. We have optimized the structure design and epitaxial fabrication process for NFAD devices to enhance the maximum achievable PDE. In addition, we have implemented a specific readout circuit and a compact thermoacoustic cryocooler to suppress the afterpulsing effect and to optimize the overall performance within a large temperature range. After performing comprehensive characterizations, the SPD exhibits an excellent overall performance of 40\% PDE, 2.3 kcps DCR, 8\% $P_{ap}$ and 49 ps FWHM timing jitter simultaneously. Such SPD could substantially improve the performance in practical applications requiring asynchronous single-photon detection in the near-infrared.

\section*{Acknowledgment}
This work is supported by the Innovation Program for Quantum Science and Technology (2021ZD0300804) and the National Natural Science Foundation of China (62175227).

\ifCLASSOPTIONcaptionsoff
  \newpage
\fi

%


\begin{IEEEbiographynophoto}{Qi Xu}
received the B.S. degree in applied physics from Anhui University, Hefei, in 2018. She is currently a Ph.D. student majoring in quantum information and quantum physics in University of Science and Technology of China. Her research interests include single-photon detection and quantum Lidar.
\end{IEEEbiographynophoto}

\begin{IEEEbiographynophoto}{Chao Yu}
received the B.S. and Ph.D. degrees in physics from the University of Science and Technology of China, Hefei, in 2014 and 2020, respectively. Since 2021, he has been a postdoctoral researcher in University of Science and Technology of China. His research interests include single-photon detection, single-photon imaging, and quantum Lidar.
\end{IEEEbiographynophoto}

\begin{IEEEbiographynophoto}{Wei Chen}
received the M.S. degree from University of Electronic Science and Technology of China. He is currently an engineer in Chongqing Optoelectronics Research Institute. His research interest is the semiconductor preparation technology.
\end{IEEEbiographynophoto}

\begin{IEEEbiographynophoto}{Jianglin Zhao}
received the M.S. degree from Harbin Engineering University. He is currently an engineer in Chongqing Optoelectronics Research Institute. His research interest is the InP photodector.
\end{IEEEbiographynophoto}

\begin{IEEEbiographynophoto}{Dajian Cui}
received the B.S. and M.S. degrees from Jiangsu Normal University, and Chongqing University, in 2005 and 2008, respectively. From 2008 to 2015, he was an engineer in Chongqing Optoelectronics Research Institute. Since 2015, he has been appointed as a senior engineer in the compound semiconductor optoelectronics division of Chongqing Optoelectronics Research Institute. His research interests include the compound semiconductor device, optoelectronic devices, and semiconductor preparation technology.
\end{IEEEbiographynophoto}

\begin{IEEEbiographynophoto}{Jun Zhang}
received the B.S. and Ph.D. degrees in physics from the University of Science and Technology of China, Hefei, in 2002 and 2007, respectively. From 2007 to 2011, he was a postdoctoral researcher in the Group of Applied Physics, University of Geneva. In 2011, he joined University of Science and Technology of China. Since 2016, he has been a full professor in University of Science and Technology of China. He has authored 70 academic articles with more than 5400 citations (Google Scholar). His current research interests include quantum communication, single-photon detection, quantum randomness, and quantum Lidar.
\end{IEEEbiographynophoto}

\begin{IEEEbiographynophoto}{Jian-Wei Pan}
received the Ph.D. degree from the University of Vienna in 1999. He is currently a Professor of Physics in University of Science and Technology of China, an Academician of Chinese Academy of Sciences (CAS), and a Fellow of the World Academy of Sciences (TWAS). He serves as the Director of the CAS Center for Excellence in Quantum Information and Quantum Physics, and the Chief Scientist for the Quantum Science Satellite Project. His research fields focus on quantum foundations, quantum optics and quantum information. He has authored over 300 articles with more than 69000 citations (Google Scholar).

His awards and honors include Fresnel Prize (European Physical Society), Quantum Communications Award, First Prize of National Prize for Natural Sciences of China, Future Science Prize in Physical Sciences, Willis E. Lamb Award for Laser Science and Quantum Optics, Newcomb Cleveland Prize (American Association for the Advancement of Science), R. W. Wood Prize (Optical Society of America), Micius Quantum Prize, and Zeiss Research Award. He was selected by Nature as ``people of the year'' in 2017 who ``took quantum communication to space and back''.
\end{IEEEbiographynophoto}







\begin{thebibliography}{1}

\bibitem{SPD11}
M. D. Eisaman, J. Fan, A. Migdall, and S. V. Polyakov, Rev. Sci. Instrum. \textbf{82}, 071101 (2011).

\bibitem{JMH15}
J. Zhang, M. A. Itzler, H. Zbinden, and J.-W. Pan, Light Sci. Appl. \textbf{4}, e286 (2015).

\bibitem{JTK00}
J. Rarity, T. Wall, K. Ridley, P. Owens, and P. Tapster, Appl. Opt. \textbf{39}, 6746-6753 (2000).

\bibitem{MCJ08}
M. Liu, C. Hu, J. C. Campbell, Z. Pan, and M. M. Tashima, IEEE J. Quantum Electron. \textbf{44}, 430-434 (2008).

\bibitem{RMG09}
R. E. Warburton, M. Itzler, and G. S. Buller, Appl. Phys. Lett. \textbf{94}, 071116 (2009).

\bibitem{CMX10}
C. Hu, M. Liu , X. Zheng, and J. Campbell, IEEE J. Quantum Electron. \textbf{46}, 35-39 (2010).

\bibitem{RDJ07}
R. T. Thew, D. Stucki, J.-D. Gautier, and H. Zbinden, Appl. Phys. Lett. \textbf{91}, 201114 (2007).

\bibitem{JRJ09}
J. Zhang, R. Thew, J.-D. Gautier, N. Gisin, and H. Zbinden, IEEE J. Quantum Electron. \textbf{45}, 792-799 (2009).

\bibitem{Xu20}
Y. Xu , J. Lu, and Z. Wu, IEEE Photon. J. \textbf{12}, 6803208 (2020).


\bibitem{XMB09}
X. Jiang, M. A. Itzler, B. Nyman, and K. Slomkowski, Proc. SPIE \textbf{7320}, 732011 (2009).

\bibitem{MXB10}
M. A. Itzler, X. Jiang, B. M. Onat, and K. Slomkowski, Proc. SPIE \textbf{7608}, 760829 (2010).

\bibitem{XMK11}
X. Jiang, M. A. Itzler, K. O'Donnell, M. Entwistle, and K. Slomkowski, Proc. SPIE \textbf{8033}, 80330K (2011).

\bibitem{FMA13}
F. Acerbi, M. Anti, A. Tosi, and F. Zappa, IEEE Photon. J. \textbf{5}, 6800209 (2013).

\bibitem{YWT20}
Y. Q. Fang, W. Chen, T. Ao, C. Liu, L. Wang, X. Gao, J. Zhang, and J.-W. Pan, Rev. Sci. Instrum. \textbf{91}, 083102 (2020).

\bibitem{Tada20}
A. Tada, N. Namekata, and S. Inoue, Jpn. J. Appl. Phys. \textbf{59}, 072004 (2020).

\bibitem{Tosi21}
F. Signorelli, F. Telesca; E. Conca, A. Della Frera, A. Ruggeri, A. Giudice, and A. Tosi, in IEEE International Electron Devices Meeting (IEDM) (2021).

\bibitem{He22}
T. T. He, X. H. Yang, Y. S. Tang, R. Wang, and Y. J. Liu, J. Semicond. \textbf{43}, 102301 (2022).

\bibitem{ZDA12}
Z. Yan, D. R. Hamel, A. K. Heinrichs, X. Jiang, M. A. Itzler, and T. Jennewein, Rev. Sci. Instrum. \textbf{83}, 073105 (2012).

\bibitem{TCO12}
T. Lunghi, C. Barreiro, O. Guinnard, R. Houlmann, X. Jiang, M. A. Itzler, and H. Zbinden, J. Mod. Opt. \textbf{59}, 1481-1488 (2012).

\bibitem{BNT14}
B. Korzh, N. Walenta, T. Lunghi, N. Gisin, and H. Zbinden, Appl. Phys. Lett. \textbf{104}, 081108 (2014).

\bibitem{CMH17}
C. Yu, M. Shangguan, H. Xia, J. Zhang, X. Dou, and J.-W. Pan, Opt. Express \textbf{25}, 14611-14620 (2017).

\bibitem{CJH18}
C. Yu, J. Qiu, H. Xia, X. Dou, J. Zhang, and J.-W. Pan, Rev. Sci. Instrum. \textbf{89}, 103106 (2018).

\bibitem{JBL16}
J. Ma, B. Bai, L.-J. Wang, C.-Z. Tong, G. Jin, J. Zhang, and J.-W. Pan, Appl. Opt. \textbf{55}, 7497 (2016).

\bibitem{Jiang18}
W.-H. Jiang, X.-J. Gao, Y.-Q. Fang, J.-H. Liu, Y. Zhou, L.-Q. Jiang, W. Chen, G. Jin, J. Zhang, and J.-W. Pan, Rev. Sci. Instrum. \textbf{89}, 123104 (2018).

\bibitem{VRG11}
V. S. Chakravarthy, R. K. Shah, and G. Venkatarathnam, J. Therm. Sci. Eng. Appl. \textbf{3}, 020801 (2011).

\bibitem{CJB11}
C. Hu, X. Zheng, J. C. Campbell, B. M. Onat, X. Jiang, and M. A. Itzler, J. Mod. Opt. \textbf{58}, 201-209 (2011).


\bibitem{MXM12}
M. A. Itzler, X. D. Jiang, and M. Entwistle, J. Mod. Opt. \textbf{59}, 1472-1480 (2012).

\bibitem{KLK15}
B. Korzh, T. Lunghi, K. Kuzmenko, G. Boso, and H. Zbinden, J. Mod. Opt. \textbf{62}, 1151-1157 (2015).

\bibitem{EGB16}
E. Amri, G. Boso, B. Korzh, and H. Zbinden, Opt. Lett. \textbf{41}, 24 (2016).

\bibitem{JYY21}
J. Liu, Y. Xu, Y. Li, Z. Liu, and X. Zhao, Opt. Express \textbf{29}, 7 (2021).

\bibitem{QCD600B}
http://www.quantum-info.com/product/hexinzujian/671.html

\bibitem{idqube}
https://www.idquantique.com/quantum-sensing/products/id-qube-nir-free-running/

\bibitem{pdmir}
http://www.micro-photon-devices.com/Products/Photon-Counters/PDM-IR

\bibitem{aurea}
http://aureatechnology.com/en/products/nir-photon-counter.html

\bibitem{Wooriro}
http://www.wooriro.com/en/bbs/board.php?bo\_table=product04\&wr\_id=4

\end{thebibliography}
\end{document}